\documentstyle[amstex,newlfont,amssymb,aps,prl,psfig]{revtex}

\begin{document}
\author{Beno\^{\i}t Gr\'emaud$^{\dagger}$ and 
Sudhir R. Jain$^{\dagger\dagger}$}
\address{$^{\dagger}$ Laboratoire Kastler Brossel, Universit\'e Pierre et
Marie Curie, T12, E1 \\
4, place Jussieu, 75252 Paris Cedex 05, France \\
$^{\dagger\dagger}$Theoretical Physics Division,
Bhabha Atomic Research Centre, \\
Central Complex, Trombay, Mumbai 400 085, India}
\title{Spacing distributions for rhombus billiards}
\date{\today}
\maketitle
\begin{abstract}
We show that the spacing distributions of rational rhombus billiards 
fall in a family of universality classes distinctly different from the 
Wigner-Dyson family of random matrix theory and the Poisson distribution.
Some of the distributions find explanation in a recent work 
of Bogomolny, Gerland, and Schmit. For the irrational billiards, despite 
ergodicity, we get the same  distribution for the examples considered -  
once again, distinct from the Wigner-Dyson distributions. All the results 
are obtained numerically by a method that allows us to reach very 
high energies.

\end{abstract}

\pacs{PACS number(s):05.45.+b }

	Statistical analysis of level correlations of a quantum system
is one of the many ways to study the effects of  chaotic behaviour of its
classical counterpart \cite{bohigas}. For such complex systems,
the fluctuations are very
well described by the random matrix theory, giving rise to three classes of
universality corresponding to orthogonal, unitary and symplectic
ensembles (OE, UE and SE).
On the other hand, for integrable systems,
the short range correlations  follow the Poisson distribution. Rhombus
billiards \cite{jain} are peculiar as they are
pseudo-integrable systems and for this
reason, their statistical properties belong to another class of
universality \cite{parab-jain}. These non-integrable systems are
termed pseudo-integrable as the dynamics occurs on a multiply-connected,
compact surface in the phase space. For example, in the case of
$\pi /3$-rhombus billiard, the invariant integral surface is a sphere
with two handles \cite{jain,jain-parab}. 
It has been  shown that the short-range properties 
(spacing distribution) can be fitted by  Brody distributions~\cite{Brody73}
with parameters depending on the genus~\cite{Shudo93}. However, a very
small number of levels were used to achieve the statistics and, as it was
outlined by the authors, the parameters were smoothly changing with the
number of levels considered. This last effect is probably a consequence of
the pseudo-integrability and thus one has to consider levels lying very high in 
energy to have converged statistics. Furthermore,  Brody distributions are
not very convenient for two reasons: (i) they are not on a firm 
theoretical basis like random matrix theory and so, one can
not gain too much knowledge about the system from the  Brody parameter; 
(ii) their behaviour at small spacing is not linear, whereas it is
so for rhombus billiards. On
the contrary, in a  recent paper~\cite{Bogomolny98}, Bogomolny et al. have 
proposed a model derived from the Dyson's stochastic
Coulomb gas one~\cite{Mehta,Haake}~: eigenvalues are considered as
classical  particles on a line, with a two-body interaction potential
given by $V(x)=-\ln(x)$. Contrary to Dyson's model, where all possible
pairs are considered, the same interaction is restricted only 
to nearest-neighbours. Hereafter, this model will be referred as 
short ranged Dyson's model (SRDM). The joint probability obtained gives rise to 
spacing distributions showing linear level repulsion and exponential
decrease for large spacing. 
More precisely, the nearest-neighbour (NN) and 
next-nearest-neighbour (NNN) distributions are~:

\begin{equation}
\label{fnsdis}
P(s)=4se^{-2s} \qquad \mathrm{and} \qquad P_2(s)=\frac 83s^3e^{-2s}
\end{equation}

\noindent 
It is worth noting that exactly the same functional form was used in 
the past \cite{date-jain-murthy} to explain the intermediate spacing 
distribution for a rectangle billiard with a flux line - an 
Aharanov-Bohm billiard.
In this recent work \cite{Bogomolny98}, it is also shown that the 
level statistics of some
rhombus billiards agree very well with these distributions. However, only
rhombi with rational angles and with Dirichlet 
boundary conditions on both $x$ and
$y$ axis (i.e. right-angled  triangle) were studied. In this Letter, we
extend the preceding study to rational billiards with Neumann boundary 
conditions (i.e. ``pure" rhombus) and also to irrational billiards (both
classes of boundary conditions). Of course, in a rhombus, making the   
shorter (longer) diagonal Neumann means that one is considering a 
larger obtuse (acute) triangle. So, the modifications are expected 
but here they are non-trivial.

The spectral properties of these systems which are non-integrable and yet 
non-chaotic is thus an important unsettled problem. The solution of this 
problem is partly in devising numerical techniques that allow one to go 
to higher energies, and, partly in developing statistical models like the 
SRDM (Ref. \cite{Bogomolny98})mentioned above. In this Letter, we first discuss 
the method and then use the levels in high energy range to show agreements 
and disagreements with the results in~\cite{Bogomolny98}. To give an idea, 
the efficiency of the method is such that we were able to compute a very 
large number of levels
(up to 36000 for a given rhombus and a given symmetry class), 
so that the statistical properties
 are fully converged. In the later part of the Letter, we show the
effects of both, the boundary conditions and the irrationality on the level
spacing distributions.

\begin{figure}
\centerline{\psfig{figure=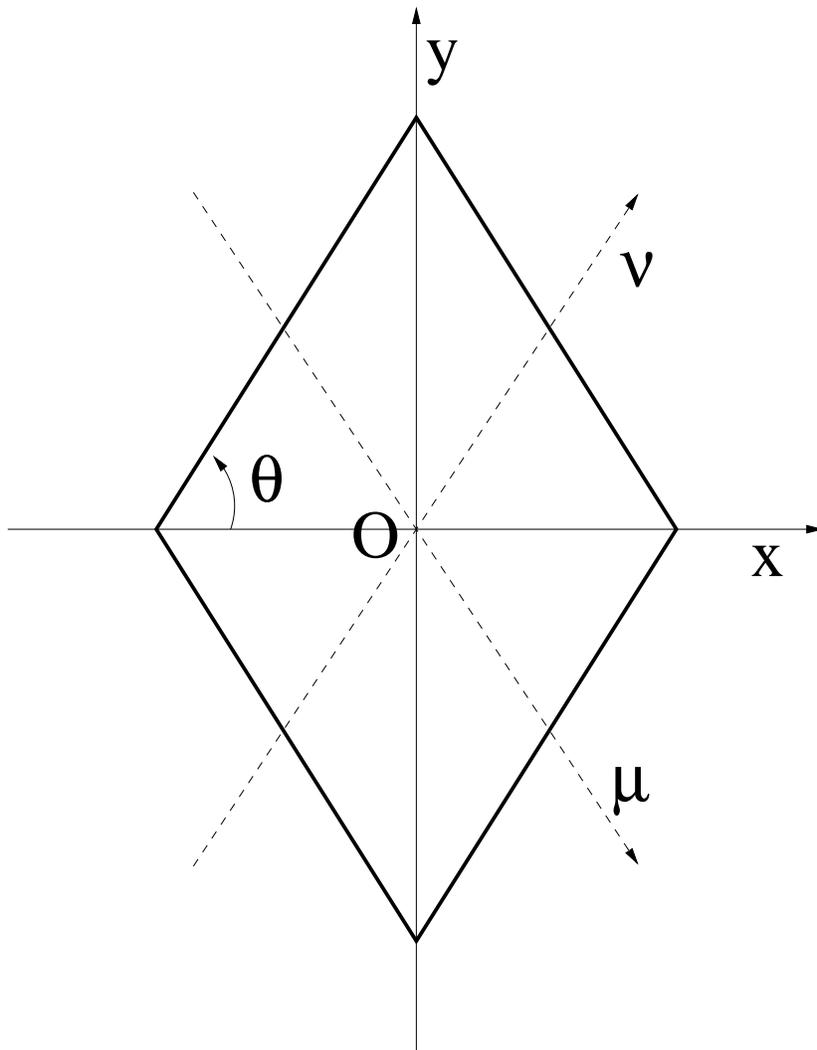,height=14cm,angle=-90}}
\caption{\label{billiard}Rhombus-shaped enclosure in which the 
particle moves freely with elastic bounces on the boundary, the quantum problem 
corresponds to imposing the 
Dirichlet conditions for the wavefunctions. The system being symmetric under
reflections with respect to the $x-$axis or the $y-$axis, Dirichlet or
Neumann boundary conditions can be imposed on the both axis, leading to four
different classes of symmetry. By  considering axes crossing at the 
centre $O$ of the system and parallel to the edges of the billiard, a
non-orthogonal coordinate system ($\mu,\nu$) 
is constructed in which the Dirichlet
boundary conditions on the enclosure separate (see text).}
\end{figure}

	The Schr\"odinger equation for a particle moving freely in a rhombus
billiard (shown by Fig.~\ref{billiard}) is simply~:

\begin{equation}
-\frac {\hbar^2}{2m}\left(\frac{\partial^2}{\partial x^2}+
\frac{\partial^2}{\partial y^2}\right)\psi(x,y)=E\psi(x,y)
\end{equation}
	
\noindent with the additional condition that $\psi(x,y)$ is
vanishing on the boundary (Dirichlet conditions). The geometry
of the system leads to a natural change of coordinates~: the two new axes
cross at the centre and are parallel to the edges of the billiard (see
Fig.~\ref{billiard})~:

\begin{equation}
\left\{
\begin{split}
\mu &=\frac 12\left(\frac x{\cos\theta}-\frac y{\sin\theta}\right); ~~ \\
\nu &=\frac 12\left(\frac x{\cos\theta}+\frac y{\sin\theta}\right).
\end{split}
\right.
\end{equation}
 
	In this new coordinate system, the original rhombus is mapped onto
a square of length $L$ and thus, in this coordinate system, the boundary conditions separate, of course at the price of a slightly more 
complicated Schr\"odinger equation~:

\begin{equation}
-\frac {\hbar^2\left(\partial^2_{\mu\mu}+
\partial^2_{\nu\nu}-2\cos(2\theta)\partial^2_{\mu\nu}\right)}{2m\sin^2(2\theta)}
\psi(\mu,\nu)=E\psi(\mu,\nu).
\end{equation}

\noindent The change $\mu\rightarrow\frac 2L\mu$, 
$\nu\rightarrow\frac 2L\nu$
and $E\rightarrow (\frac 2L)^2\frac m{\hbar^2}E$ gives rise to the scaled
Schr\"odinger equation (after multiplication by $2\sin^2(2\theta)$)~:

\begin{equation}
\label{schrodinger}
-\left(\partial^2_{\mu\mu}+
\partial^2_{\nu\nu}-2\cos(2\theta)\partial^2_{\mu\nu}\right)
\psi=2\sin^2(2\theta) E\psi,
\end{equation}

\noindent the boundary condition being then at the points $\mu=\pm1$ and
$\nu=\pm1$.

	To solve the eigenvalue problem, a possible idea is to expand any
wavefunction in a basis satisfying the boundary 
conditions:

\begin{equation}
\psi(\mu,\nu)=\sum_{n_{\mu},n_{\nu}=0}^{\infty}a(n_{\mu},n_{\nu})
\phi_{n_{\mu}}(\mu)\phi_{n_{\nu}}(\nu)
\end{equation}

\noindent The simplest choice is the Fourier sine and cosine series.
Unfortunately, the operator 
$\partial^2_{\mu\nu}$ has no selection rules in this
basis, thus the matrix representation of the left part of the
Schr\"odinger equation ~\eqref{schrodinger} is totally filled. Numerically,
we will approximate the wavefunction by keeping only a (large) number
of terms in the preceding serie. For this system and for many 
others like coulomb systems, it has
been observed that the rate of convergence of the serie is much slower
when the matrix is filled than when selection rules occur.

	To avoid this difficulty,  we introduce the following basis for each
coordinate $\mu$ and $\nu$~:

\begin{equation}
\phi_n(u)=(1-u^2)C_n^{(\frac 32)}(u)
\end{equation}

\noindent where $C_n^{\alpha}$ are Gegenbauer polynomials~\cite{Abramovitz}. 
This basis is complete and all operators appearing in 
equation ~\eqref{schrodinger} have selection rules. More precisely, we have

\begin{equation}
|\Delta n_{\mu}|,|\Delta n_{\nu}|\le 2\qquad 
\Delta n_{\mu}+\Delta n_{\nu}=0,\pm2,\pm4
\end{equation}

\noindent Furthermore, all matrix elements are analytically known and are
given by simple polynomial expressions of the two quantum numbers
$(n_{\mu},n_{\nu})$. The only difficulty is the non-orthogonality of the
basis, that is $\langle n'|n\rangle$ does not reduce to $\delta_{nn'}$, but
also shows the preceding selection rules. 

	This basis also allows us to take directly into account the 
symmetries of the original problem, namely the reflections with respect to
the $x$-axis ($S_x$) or the $y$-axis ($S_y$). In $(\mu,\nu)$ coordinates, 
they
become~:

\begin{equation}
S_x\quad\left\{
\begin{split}
\mu&\rightarrow\nu ,\\
\nu&\rightarrow\mu
\end{split}\right. ;
\qquad\qquad S_y\quad\left\{
\begin{split}
\mu&\rightarrow-\nu ,\\
\nu&\rightarrow-\mu. 
\end{split}\right.
\end{equation}

\noindent Using the properties of the Gegenbauer 
polynomials~\cite{Abramovitz}, we are able to construct  four 
different bases in which the two operators $S_x$ and $S_y$ are 
simultaneously diagonal with eigenvalues, 
$\epsilon_x=\pm1$ and $\epsilon_y=\pm1$. Of course, this transformation 
preserves the selection rules and hence  the band structure. 
We shall denote the eigenfunctions vanishing on both the diagonals by 
($--$)- and not vanishing on either by ($++$)-parity classes. 

	The original Schr\"odinger equation is thus transformed to a
generalized eigenvalue problem~:

\begin{equation}
A|\psi\rangle=EB|\psi\rangle
\end{equation}

\noindent where $A$ and $B$ are real, sparse and banded matrices. This kind of
system is easily solved using the Lanczos algorithm~\cite{lanczos}. 
It is an iterative method, 
highly efficient to obtain few eigenvalues and eigenvectors 
of very large matrices. We
obtain typically 100 eigenvalues of a $10000\times10000$ matrix in few 
minutes on a regular workstation. The results presented here have been
obtained by diagonalizing matrices of size up to $203401$ for a bandwidth
equal to $903$. For such matrices, we obtain 200 eigenvalues in 10 minutes
on a Cray C98. The number of levels ($\simeq 36000)$ that we are able to
compute in this way is a bit larger than with usual boundary matching methods
$(\simeq 20000)$, which are nevertheless restricted to rational angles. On the
other hand, very recent methods developed by
Vergini et al~\cite{Vergini95}  seems to be more efficient (they were able to
reach energy domain around the 142,000$^{\mathrm{th}}$ state for the stadium 
billiard).

\begin{figure}
\centerline{\psfig{figure=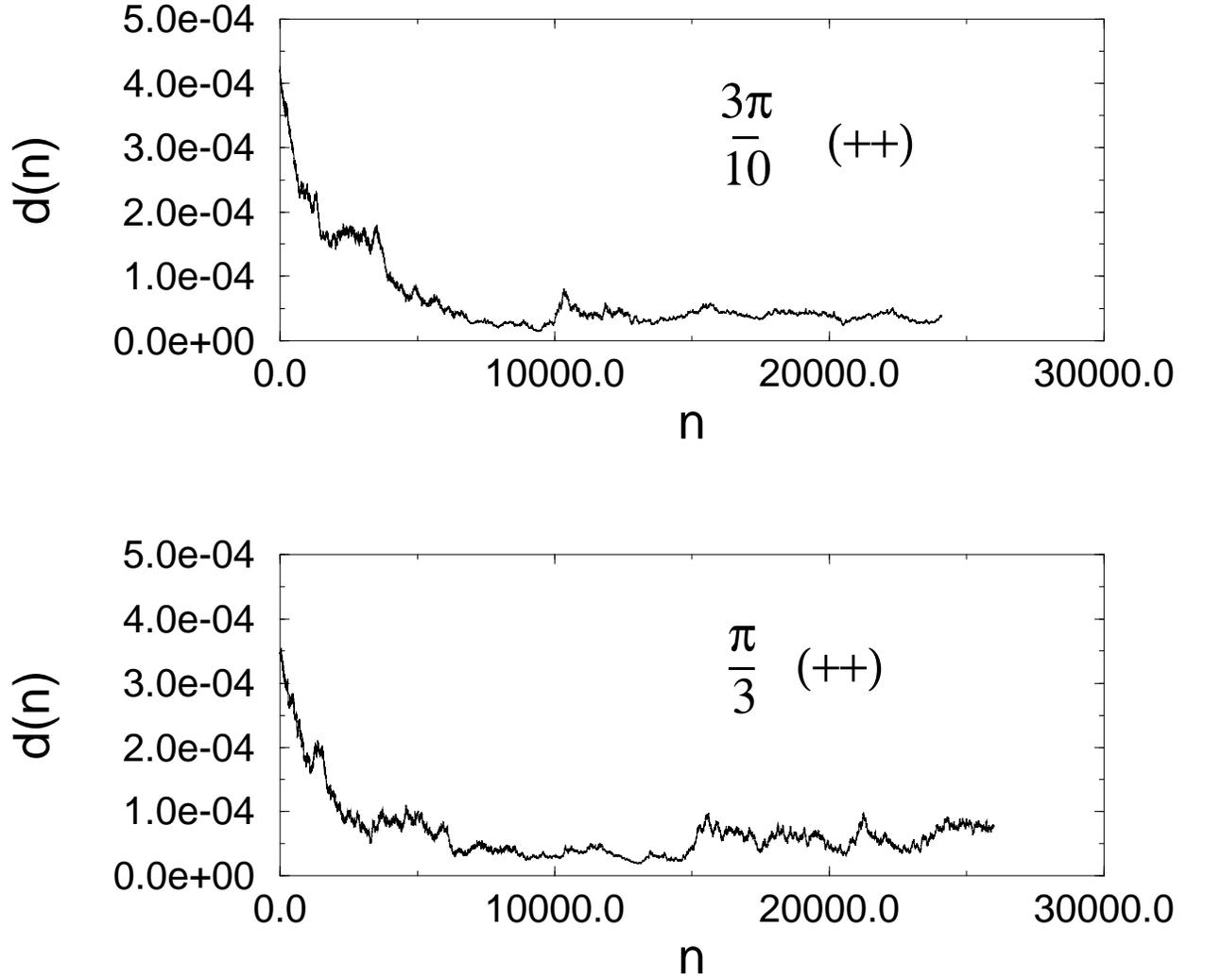,height=14cm,angle=-90}}
\vspace{2cm}
\caption{\label{converg} ``Difference'' (see equation~\eqref{dist}) 
between the NN statistics obtained
with the 5000 highest states and the NN statistics obtained with levels
$n$ to $n+4999$, as a function of $n$, for both $\frac{3\pi}{10}$ (top) and 
$\frac{\pi}3$ (bottom) ($++$ parity). Above the 10000$^\mathrm{th}$ level,
the distributions become energy  independent (apart fluctuations).}
\end{figure}

	For the present study, various 
values of angle have been used~: 
\begin{equation}
\frac{3\pi}{10},\quad \frac{(\sqrt{5}-1)\pi}4,\quad \frac {\pi}{\pi},\quad
\frac{\pi}3\quad \frac{3\pi}8 \quad \mathrm{and}\quad\frac{7\pi}{18}
\end{equation}
\noindent for both $(++)$ and $(--)$ parity~\cite{pis3}. For all cases, only
levels above the $10000^{\mathrm{th}}$ one have been considered, to avoid
peculiar effects in the statistics and at least
5000 levels (up to 24000) have been used for each case.
The convergence of the statistics has been checked by
systematically varying the energy around which levels were taken. This is
shown in Fig.~\ref{converg}, where we have plotted the following quantity~:

\begin{equation}
\label{dist}
\int_0^{\infty}ds\,\left(N_0(s)-N_n(s)\right)^2
\end{equation}
\noindent with respect to the number $n$, for $\frac{3\pi}{10}$ 
(top) and $\frac{\pi}3$ (bottom) billiards ($(++)$ parity). 
$N_0(s)$ is the cumulative NN
distribution obtained with the 5000 highest states, whereas $N_n(s)$ is the
cumulative NN distribution obtained with levels $n$ to $n+4999$. One can thus
clearly see that the statistics become energy independent (up to fluctuations) 
only for levels above the 10000$^{\mathrm{th}}$ state, which emphasizes the
choice of keeping only those states.

\begin{figure}
\centerline{\psfig{figure=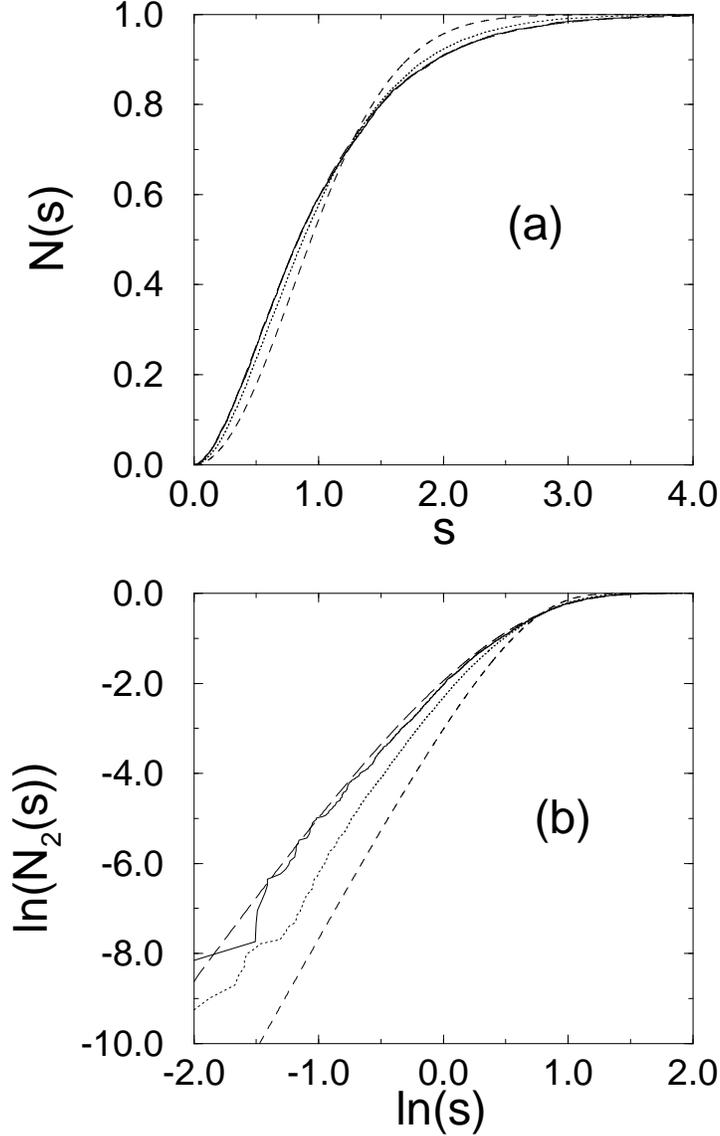,height=16cm,angle=180}}
\caption{\label{fig2}On the top is shown the cumulative distribution of 
nearest neighbour spacings for the $\frac{3\pi}{10}$ rhombus. The dotted
line corresponds to Neumann-Neumann $(++)$ boundary conditions on both $x-$ and 
$y-$axis, the continuous line corresponding to Dirichlet-Dirichlet $(--)$
boundary conditions. The two distributions are clearly different, the deviation
being larger than statistical fluctuations. The $(--)$ symmetry class is
exactly on the top of the distribution introduced by Bogomolny et al. (SRDM)
given by equation~\eqref{fnsdis},
corresponding to the long-dashed line. The $(++)$ distribution lies in
between SRDM and OE prediction (given by the dashed line). This difference is
emphasized in figure (b) depicting the next-nearest neighbour distributions
(cumulative) for the same billiards ($\ln-\ln$ plot). 
Again, the $(--)$ (continuous line) symmetry class is
exactly on the top of SRDM (long dashed line), whereas the $(++)$ symmetry class
(dotted line) lies in between SRDM and OE (dashed line). Especially the
behaviours for small spacing are very different~: $(--)$ shows a $s^4$
power law, whereas it is $s^5$ for $(++)$, the OE prediction.}  
\end{figure}

\begin{figure}
\centerline{\psfig{figure=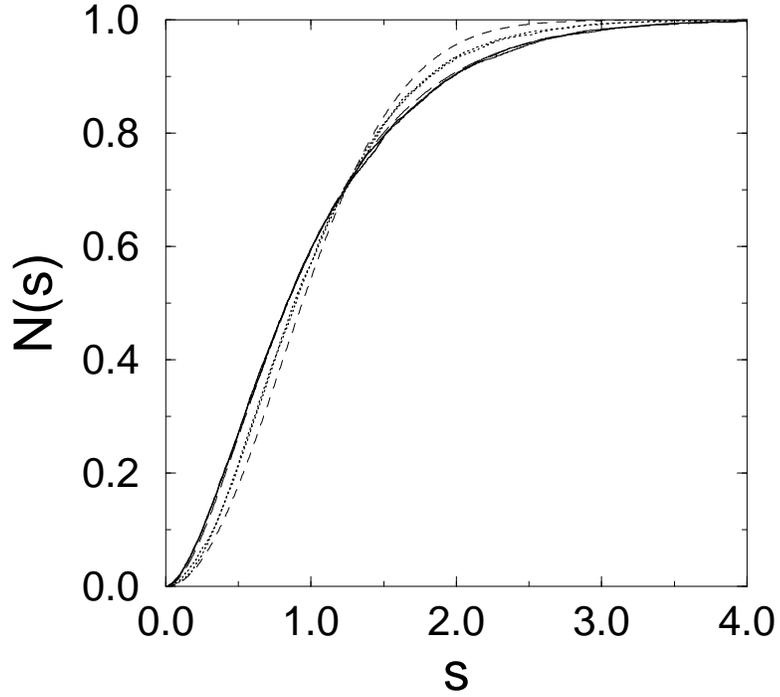,height=14cm,angle=-90}}
\vspace{1cm}
\caption{\label{fig3}Spacing distributions (cumulative) for two rational
billiards~: $\frac{3\pi}{8}$ (continuous lines) and $\frac{7\pi}{18}$ (dotted
lines), for both $(++)$ and $(--)$ symmetry classes. Contrary to the
$\frac{3\pi}{10}$ billiard (see Fig.~\ref{fig2}), there is no difference 
between the two symmetry classes~: for each billiard the two curves lie on
 top of each other. Furthermore, these two billiards show distinct
spacing distributions, the $\frac{3\pi}{8}$ one corresponds exactly to SRDM
(long dashed line) whereas the $\frac{7\pi}{18}$ one is much closer to OE
prediction (dashed line).}
\end{figure}

	In~\cite{Bogomolny98}, it was shown that for the 
$\frac{3\pi}{10}(--)$
billiard, both NN and NNN statistics were following the 
formula~\eqref{fnsdis}.
We, of course, reproduce this result, as shown in Fig.~\ref{fig2}(a). However,
the same billiard, but with Neumann-Neumann boundary conditions does not
follow the same distribution laws, but rather lies in between OE and 
SRDM distributions, as it is shown in Fig.~\ref{fig2}(a).
The deviations are obviously much larger than statistical fluctuations. The
difference is emphasized by looking at the behavior of the NNN for small
spacings (see Fig.~\ref{fig2}(b)). Indeed, whereas for the $(--)$ symmetry, the
observed power law is $s^4$ in the cumulative distribution (i.e.,  $s^3$ for
$P(1,s)$), it is close to $s^5$ for the $(++)$ case (i.e. $s^4$ for
$P(1,s)$), which is the OE prediction. This dependency of the statistics
with respect to the boundary conditions has
already been observed in other systems like  3D Anderson's 
model~\cite{Montambaux97}. However the present results are more surprising
as there are $\theta$ values for which there is practically  
no difference between the
two symmetry classes. Indeed, Fig.~\ref{fig3} shows the NN (cumulative)
distributions for $\frac{3\pi}8$ and $\frac{7\pi}{18}$. 
Besides the statistical fluctuations, one cannot distinguish between the
two symmetry classes, whereas the distributions differ~:$\frac{3\pi}8$ is
well described by SRDM, whereas $\frac{7\pi}{18}$ lies
between OE and SRDM. 

The case of the $\frac{\pi}3$
billiard is the most peculiar, since the $(--)$ parity is integrable whereas
the $(++)$ spacing distributions agree with SRDM. 

All the rhombi considered are not ergodic, as their genus
are finite (e.g., two for the $\pi /3$-rhombus). 
In contrast, for irrational angle, ``the genus is infinite", and
so one could expect a rather different behaviour. 
Although the concept of genus is applicable only to compact surfaces, we 
have stated the above phrase in quotes in the following sense. 
As an irrational rhombus is approximated via continued fraction expansion, 
the larger and larger denominators will appear, implying larger genus 
surfaces, and eventually ``infinite". It is quite possible, and it may, 
in fact, be  true, that this limit is singular. As a result, from the 
rational convergents, it may not be possible to say something about the 
irrational billiard.

\begin{figure}
\centerline{\psfig{figure=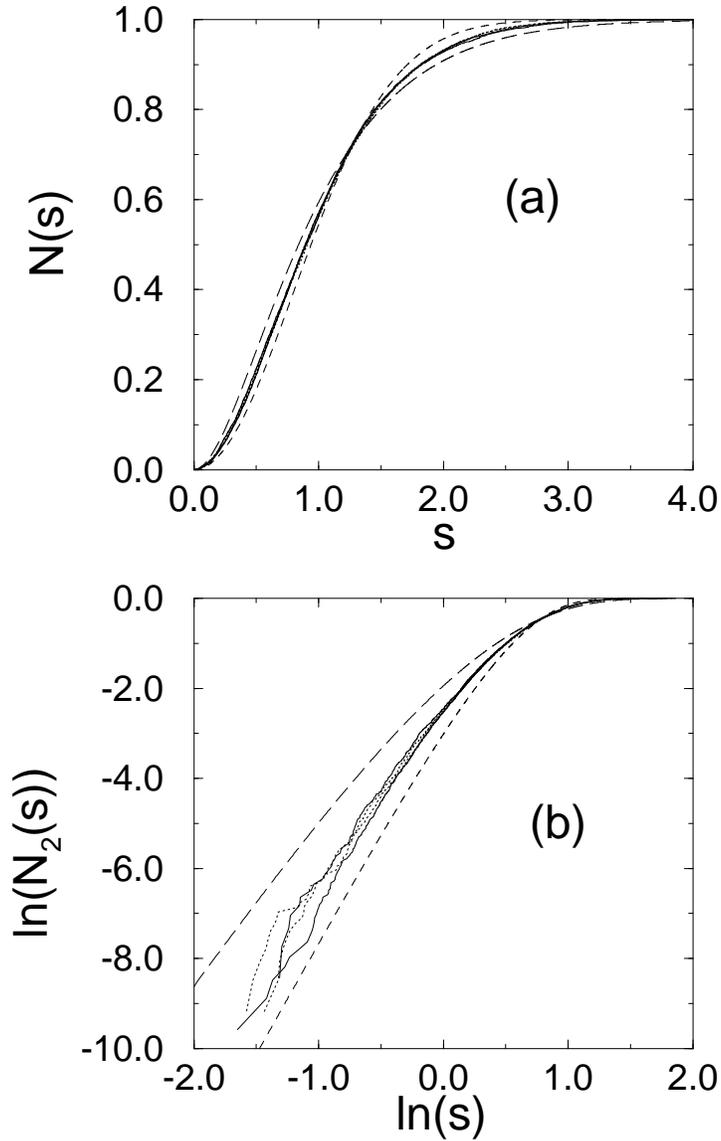,height=16cm,angle=180}}
\caption{\label{fig4}Nearest neighbour (a) and next-nearest neighbour (b)
distributions for two irrational billiards~: $\frac{\pi}{\pi}$ (continuous
lines) and $\frac {(\sqrt{5}-1)\pi}{4}$ (dotted lines) for both
$(++)$ and $(--)$ symmetry classes. Contrary to the rational billiards, the
genus of these billiard is ``infinite" (see text for explanation) and so the
classical dynamics is ergodic. The fact that all the four distributions 
lie on
top  of each other is quite remarkable and may be related to
the fact that these billiards have the ``same" genus. However, from the
ergodicity one could expect the distributions to be OE-like, which is not
the case. They rather lie between SRDM (long dashed) and OE (short dashed).
Still, the small spacing behaviour of the next-nearest neighbour
distributions shows a $s^5$ power law, i.e. corresponding to OE.}
\end{figure}

Fig.~\ref{fig4} displays
NN and NNN statistics for $\frac{\pi}{\pi}$ and $\frac{(\sqrt{5}-1)\pi}4$
(both symmetry classes) billiards. NN distributions are on  top  of each other,
which is interesting if one believes that the genus is the 
relevant parameter. On
the other hand, from the ergodicity, one could expect the distributions to be
OE, which is not the case, even if the small spacing behavior of NNN
statistics seems to show the same power law $s^5$ (for cumulative).
Thus, if Ref.\cite{Bogomolny98}  seems to give one class of
universality, there must be other classes of universality lying between SRDM
and OE, especially for irrational angles. The other possibility is that,
although numerically stationary in a wide range of energy, the spacing 
distributions of irrational rhombus may evolve exceptionally slowly to OE.
If it is the case, one will probably have to find the final answer in a much
higher energy range, for which other numerical methods will have to be 
used~\cite{Vergini95}. 

The present study also raises the question of the
semiclassical understanding of the boundary dependence of 
the distributions. Due to change in boundary conditions, 
actions, Maslov indices and also the edge orbits 
will change resulting in a difference, but the whole explanation of 
this boundary dependence probably lies beyond these simple considerations.
Spectral fluctuations in some of the pseudo-integrable 
billiards have been studied in detail using the periodic orbit theory. 
From the detailed information about the periodic orbits \cite{jain-parab}, 
it was shown that the spectral rigidity is non-universal \cite{parab-jain} 
with a universal trend. 
We hope that the method presented here, and the ensuing 
numerical results will help us to model the spectral fluctuations of these 
apparently simple non-integrable quantum systems. 

To summarize, we have casted the problem of a particle in rhombus-shaped 
enclosure in a way that allows us to go to very high energies. This has 
led us to confidently obtain statistical results on spacing distributions 
which are well converged. Subsequently, we have shown that for some rational 
billiards, the fluctuations agree well with the results recently obtained 
\cite{Bogomolny98}. However, we have given several examples where the 
recent model does not explain the obtained distributions. It has been 
shown that boundary conditions play an important role. Finally, for the 
irrational rhombus billiards, the distributions seem to be identical for the 
examples considered. Significantly though, the distribution is still not in the 
Wigner-Dyson family. We believe that these results point in the direction 
of having a family of universality classes which, in essence, leads 
to nonuniversality with a universal trend for pseudointegrable billiards. 

	We acknowledge stimulating discussions with D.~Delande and 
E.~Bogomolny.
CPU time on a Cray C98 computer has been provided by IDRIS.
Laboratoire Kastler Brossel is laboratoire de l'Universit\'e Pierre et Marie
Curie et de l'Ecole Normale Sup\'erieure, unit\'e associ\'ee 18 du CNRS.

\begin{thebibliography}{99}
\bibitem{pis3} The $(--)$ parity for the $\frac{\pi}3$ rhombus is not
shown as it is integrable.
\end{thebibliography}
\end{document}